\documentclass[11pt,a4paper]{article} 

\usepackage{anysize}
\usepackage[format = hang]{caption}
\usepackage{amsmath,amsthm,verbatim,amssymb,amsfonts,amscd, graphicx}
\usepackage{graphics,natbib}
\usepackage{titlesec,footmisc}
\usepackage{hyperref} 

\usepackage{xcolor}
\newcommand{\rr}[1]{\textcolor{red}{\bf #1}}

\usepackage{listings}
\usepackage{booktabs} 

\font\smallrm=cmr8

\def\fermat#1{\setbox0=\vtop{\hsize4.00pc
        \smallrm\raggedright\noindent\baselineskip9pt
        \rightskip=0.5pc plus 1.5pc #1}\leavevmode
        \vadjust{\dimen0=\dp0
        \kern-\ht0\hbox{\kern-4.00pc\box0}\kern-\dimen0}}

\font\cyr=wncyr10 scaled 1095
\def\e3{\rm\"{\cyr e}}
\def\bolde3{\bf\"{\cyb e}}
\def\bigbolde3{\bigbf\"{\bigcyb e}}
\def\itae3{\it\"{\cyi e}}
\def\sse3{\ssfont\"{\cyss e}}
\def\smalle3{\smallrm\"{\smallcyr e}}
\def\j3{{\rm\u{\cyr i}}}

\def\onegin{\cyr Evgeni\j3{} Onegin}
\def\markovchains{\cyr tsepi Markova}

\theoremstyle{plain}

\theoremstyle{definition}

\def\tr{{\rm t}}
\def\midd{\,|\,}
\def\pr{{\rm pr}}

\def\half{\hbox{$1\over 2$}}

\def\hopp{\medskip\noindent}
\def\hoppp{\bigskip\noindent}

\def\beq{\begin{eqnarray}}
\def\eeq{\end{eqnarray}}

\def\beqn{\begin{eqnarray*}}  
\def\eeqn{\end{eqnarray*}}

\def\pr{{\rm pr}}

\def\half{\hbox{$1\over2$}}

\def\midd{\,|\,}
\def\tr{{\rm t}}

\def\sort{{\rm sort}}
\def\prop{{\rm prop}}
\def\prev{{\rm prev}} 

\titleformat{\section}{\normalfont\large\sc\centering}{\thesection}{1em}{}
\titleformat{\subsection}[runin]{\normalfont\large\bfseries}{\thesubsection}{1em}{}
\numberwithin{equation}{section} 
\renewenvironment{abstract}
               {\list{}{\rightmargin\leftmargin}%
                \item[\text{\hspace{10mm}\sc Abstract.}]\relax}
               {\endlist}

\begin{document}

\begingroup
\begin{centering} 
  \Large{\bf Sudoku Solving and Finding Magic Squares \\
    by Probability Models and Markov Chains}
  \\[0.8em]
\large{\bf Nils Lid Hjort} \\[0.3em] 
\small {\sc Department of Mathematics, University of Oslo} \\[0.3em]
\small {\sc April 2026\footnote{material partly from
  two different FocuStat Blog Posts, 2019; 
  in this modified, combined, extended form April 2026 for wider channels}}\par
\end{centering}
\endgroup



\begin{abstract}
\small{
  The sudoku 
  puzzles have a long history, with variations going back
  more than a hundred years, but its current and perhaps surprising
  world-wide prominence goes back to certain initiatives and then
  puzzle-generating computer programmes from just after 2000.
  To solve a sudoko puzzle, a statistician can put up a probability
  model on the enormous space of $9\times9$ matrix possibilities,
  constructed to favour `good attempts', and then engineer
  a Markov chain to sample a long enough chain of sudoku table
  realisations from that model, until the solution is found.
  The methods work also for other types of puzzles, like
  constructing `magic squares' with wished-for properties
  (sums of rows, columns, diagonals equal, etc.),
  as is also illustrated in this article;
  via magic models and equally magic Markov chains 
  I find impressively magic $8\times8$ and $10\times10$ squares. 

 }
\end{abstract}


\begin{figure}[h]
\centering
\includegraphics[scale=0.66]{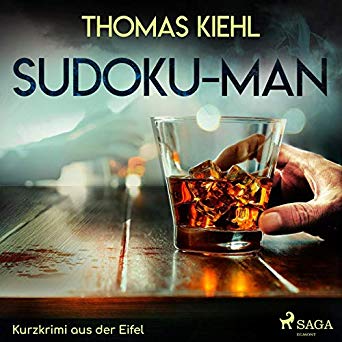}
\caption{Scary but fascinating stuff.}
\label{figure:queen1}
\end{figure}

\section*{Sudoku: the problem}
\label{section:intro}

\begin{figure}[h]
\centering
\includegraphics[scale=0.16]{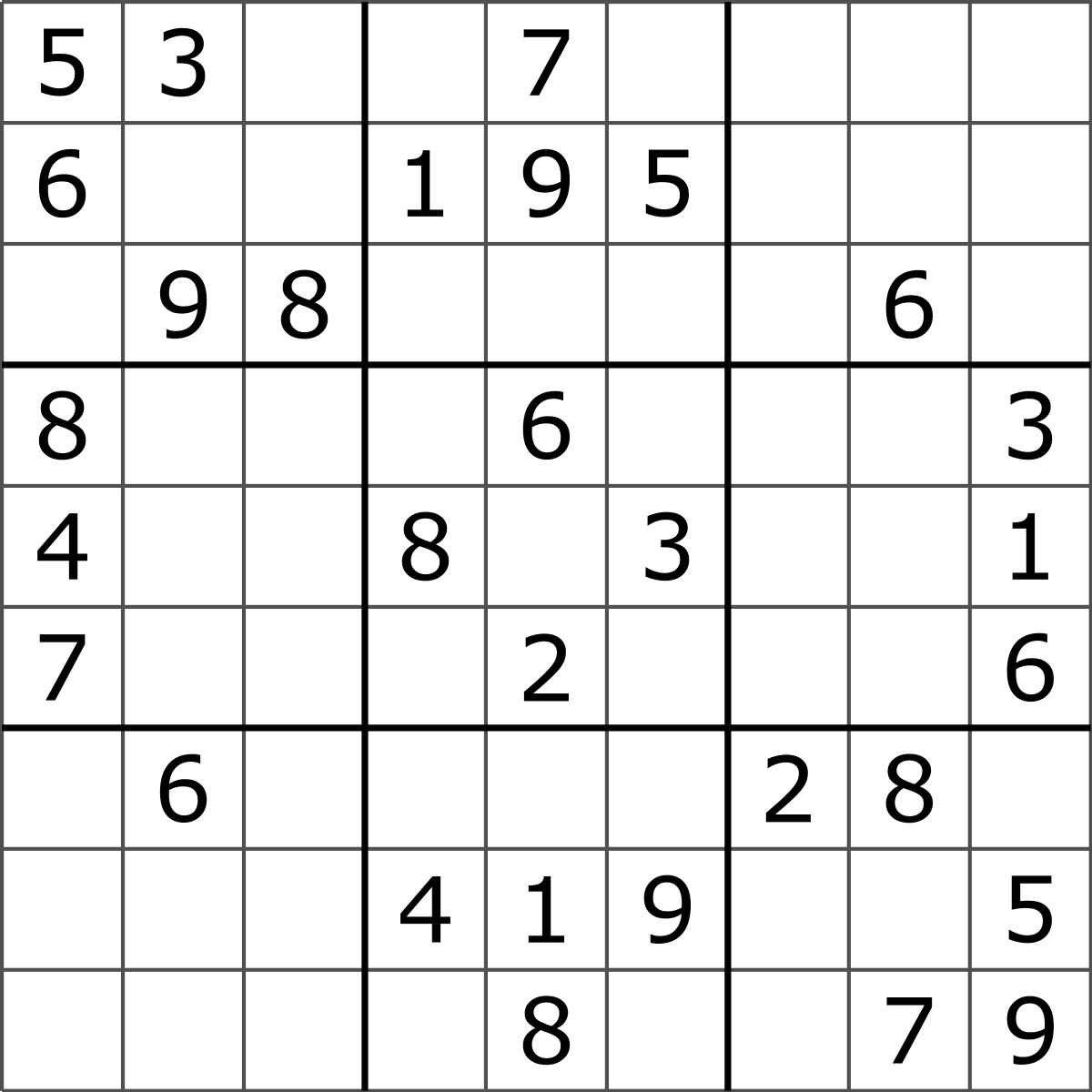}
\caption{My first-ever sudoku puzzle.}
\label{figure:queen2}
\end{figure}

Here's an example (Figure \ref{figure:queen2}),
grabbed from one of a great many sudoku
internet sites. So there are 9 blocks of size $3\times3$,
with some given numbers in given positions, but with blanks waiting
for the clever puzzler to fill in the rest. Each $3\times3$ block
should contain the numbers 1, 2, 3, 4, 5, 6, 7, 8, 9,
and the challenge is to organise these such that each row
and each column of the resulting $9\times9$ matrix contains
precisely these 1-to-9 numbers.

\section{A probabilistic-statistical take}

I've never solved a sudoku puzzle\footnote{not to be confused with
{\it sodoku}, the bacterial zoonotic disease} before, 
and to me it looks dauntingly complex (though presumably
not to millions of seasoned sudoku puzzlers, and I'm told
the one I grabbed here is not among the hard ones).
There are 4, 5, 8 missing numbers
for the first three blocks, those of the first row,
which means there are
\beqn
4!\cdot 5!\cdot 8!=24\cdot120\cdot40320=116121600 
\eeqn 
ways of organising these missing numbers alone. In total,
for the nine blocks, with missing pieces numbering
4, 5, 8, 6, 5, 8, 8, 5, 4, there's an astronomical number of
$8.388\cdot10^{23}$ different possibilities. And only one
of these is the right solution to the puzzle.

Rather than (a) going through all possibilities, which would
take my laptop longer time than the remaining life-time
of the sun and the moon; or (b) trying to learn to be
clever at these things; I choose (c) to attempt to be clever
in a different and statistical way, by building
a good enough probability model, on this enormous
sample space of $9\times9$ sudoku tables, favouring outcomes
which can't be very far from the Real Thing; and
(d) to throw a Markov chain at it, to read off those
simulated sudoku tables which have high probability
under my constructed model.

First, the sample space for my model is the colossal
one consisting of all $9\times9$ matrices $x$,
where each of the nine $3\times3$ blocks consists of the natural
numbers 1-to-9, and with the 30 fixed numbers in their
fixed positions, as given by the puzzle. So for the
upper-left corner block, I can shuffle the missing numbers
1, 2, 4, 7 around on the four available places, etc.
The size of the full sample space is the gargantuan $8.388\cdot10^{23}$ 
mentioned above. Then, my probability model, which should
favour `good attempts at solving the puzzle'. I make it
\beqn
f(x)=(1/k)\exp\{-\lambda Q(x)\}, 
\eeqn 
where $Q(x)$ is constructed to be zero for the perfect
solution and otherwise with low values for matrices that
are not too far off. Here $\lambda$ is a positive tuning
parameter, and $k$ is the normalisation constant,
a mathematical necessity, though it is impossible to compute,
since it is a sum of astrologically many terms.
As we shall see, we can do the sampling without knowing
its value, however. There are many $Q(x)$ functions
to choose from, perhaps more clever than my own brute-force
version, which is
\beqn
Q(x)=\sum_{i=1}^9 r_i(x) + \sum_{j=1}^9 c_j(x), 
\eeqn 
where
\beqn
r_i(x)&=&\sum_{j=1}^9 |\sort(x[i,\cdot])_j-j|, \\
c_j(x)&=&\sum_{i=1}^9 |\sort(x[\cdot,j])_i-i|.
\eeqn 
The $r_i(x)$ is zero if and only if row $i$
precisely contains the 1-to-9 numbers, and similarly with
the column scores $c_j(x)$. If the sudoku table $x$ 
is not far from the perfect $x^*$, the $Q(x)$ 
will be low; correspondingly, $Q(x)$ will be high
if rows and columns of $x$ are some distance away
from containing the 1-to-9 numbers. There is only one $9\times9$
sudoku table, say $x^*$, which attains $Q(x)=0$. 

\section{Markov chains}

The point now, theoretically and practically, is that I
can manage to sample sudoko tables from the $f(x)$ 
probability distribution, using an appropriately constructed
Markov chain of constrained $9\times9$ matrices. Perhaps 99.999999
percent of the umpteen fantasticatillion matrices have
very low probabilities, under my $f(x)$ model.
So if I succeed in simulating a sequence of outcomes from the model,
what I will see are $9\times9$ sudoku tables with low scores of $Q(x)$, 
i.e.~which are not far from the sudoku solution $x^*$. 

I manage this by setting up a perhaps very long Markov chain
of $9\times9$ matrices, say $x_1,x_2,x_3,\ldots$, where (a) the next
outcome $x_{t+1}$ should be easy to generate given the present $x_t$,
and (b) the equilibrium distribution should be precisely the
$f(x)$ above. 

What I do is to (i) start anywhere, e.g.~by filling in the
remaining numbers 1, 2, 4, 7 randomly in the upper-left block, etc.;
(ii) generate simple proposals, say $x_\prop$ after having seen
$x_\prev$ with `prop' and `prev' indicating the proposal and
previous sudoku table of the Markov chain; and then (iii)
accepting or rejecting these proposals with fine-calibrated
probabilities. This last step (iii) involves accepting
the proposal with probability
\beqn
\pr=\min\{1, f(x_\prop)/f(x_\prev)\}
   =\min\{1,\exp[-\lambda\{Q(x_\prop)-Q(x_\prev)\}],
\eeqn 
and letting $x_{t+1}$ remain at the previous $x_t$
with the complementary probability $1-\pr$. 
If the generated proposal has lower $Q(x_\prop)$ 
than the previous $Q(x_\prev)$, 
then it is automatically accepted as the next matrix in
my chain of attempted sudoku solutions. If it is bigger,
and hence looking worse, from the sudoku puzzle perspective,
it may still be accepted, but then with a certain
well-chosen probability, namely $\exp(-\lambda)$ 
raised to a certain power, as per the recipe described.
Note that the awkard normalisation constant $k$ 
has dropped out of the algorithm. I do step (ii)
by randomly selecting one of the nine $3\times3$ blocks,
then pointing to two of the moveable unknowns
(inside that block), and ordering these two guys to
change places. I've accomplished this in brute-force fashion,
with a decent amount of book-keeping in my R code,
and it's not at all optimised for speed -- but it works,
and can churn through a hundred thousand iterations
in three minutes on my laptop.

Theory for stationary Markov chains, with theorems from the
Kolmogorov era of the 1930ies, now guarantee that
the chain I've described has a unique equilibrium distribution,
namely \& precisely, the intended $f(x)$. 
The essence here is that with $\pi(x'\midd x)$ 
denoting the transition probability for the chain, the chance
of the next matrix being $x'$, when starting from $x$, we do have
\beqn
f(x)\pi(x'\midd x)=f(x')\pi(x\midd x')
\eeqn 
for all matrices $x,x'$ in the sudoku sample space. 

With a bit of twiddling and R-fumbling and fine-tuning for $\lambda$ 
I got the chain to work, the $Q(x)$ really visited zero
-- admittedly after quite some time, taking fifteen minutes
on my laptop, considerably more time, for this occasion,
than clever humans would need -- and found the solution,
lo {\it\&} behold:
\begin{large}
\beqn
\begingroup
\setlength{\arraycolsep}{8pt}  
\begin{matrix}
\begin{bmatrix}
  \rr{5} & \rr{3} & 4 \\
  \rr{6} & 7 & 2 \\
  1 & \rr{9} & \rr{8}
\end{bmatrix} \quad 
\begin{bmatrix}
  6 & \rr{7} & 8 \\
  \rr{1} & \rr{9} & \rr{5} \\
  3 & 4 & 2
\end{bmatrix} \quad
\begin{bmatrix}
  9 & 1 & 2 \\
  3 & 4 & 8 \\
  5 & \rr{6} & 7
\end{bmatrix} \\ \\[1pt] 
\begin{bmatrix}
  \rr{8} & 5 & 9 \\
  \rr{4} & 2 & 6 \\
  \rr{7} & 1 & 3
\end{bmatrix} \quad
\begin{bmatrix}
  7 & \rr{6} & 1 \\
  \rr{8} & 5 & \rr{3} \\
  9 & \rr{2} & 4
\end{bmatrix} \quad
\begin{bmatrix}
  4 & 2 & \rr{3} \\
  7 & 9 & \rr{1} \\
  8 & 5 & \rr{6}
\end{bmatrix} \\ \\[1pt]
\begin{bmatrix}
  9 & \rr{6} & 1 \\
  2 & 8 & 7 \\
  3 & 4 & 5
\end{bmatrix} \quad
\begin{bmatrix}
  5 & 3 & 7 \\
  \rr{4} & \rr{1} & \rr{9} \\
  2 & \rr{8} & 6
\end{bmatrix} \quad
\begin{bmatrix}
  \rr{2} & \rr{8} & 4 \\
  6 & 3 & \rr{5} \\
  1 & \rr{7} & \rr{9}
\end{bmatrix}
\end{matrix}
\endgroup 
\eeqn 
\end{large}

\begin{figure}[h]
\centering
\includegraphics[scale=0.55]{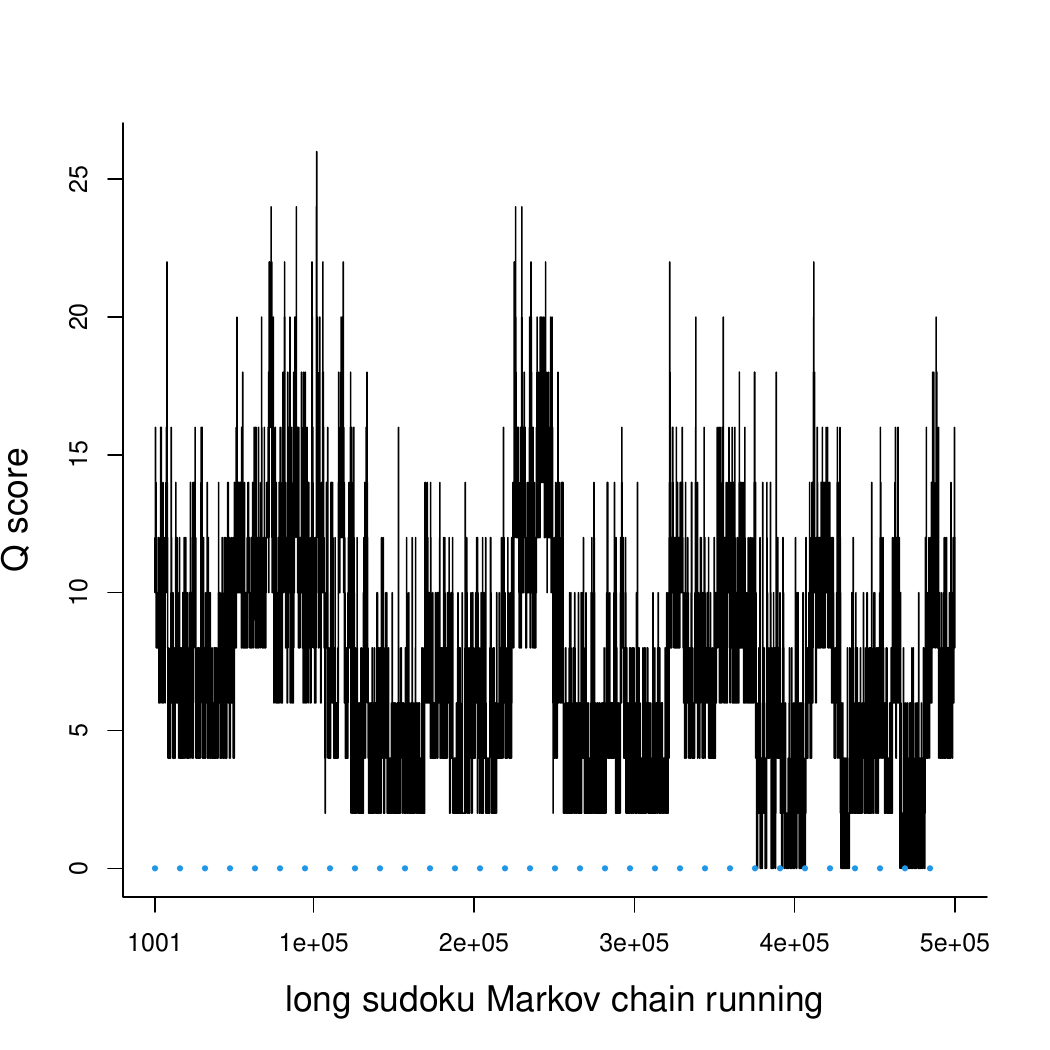}
\caption{Long sudoku Markov chain to see when $Q(x)$
  hits zero, after which I can read off the solution $x^*$.
  In the figure I have discarded the first 1000 of
  the half a million random iterations.}
\label{figure:queen3}
\end{figure}

For this particular long-running chain, shown
in Figure \ref{figure:queen3}, I used length ${\rm sim}=5\cdot10^5$,
i.e.~half a million, and $\lambda=1.333$.
I start the chain `somewhere', randomly chosen, and far away from
the solution, which means $Q(x)$ starts out high before
it comes to its stationary distribution;
in the figure, the first 1000 tables are discarded.
The $Q(x)$ reaches its implied stationary distribution
from that of the full $9\times9$ random matrix $x$. 
The very first time the chain hits zero, and the solution $x^*$
is found, happens after 376300 steps (and then the chain
finds the solution, again and again and again, 17311 more times,
before the half a million clock stops. 

\section{Finding magic squares} 

Benjamin Franklin (1706--1790), statesman, inventor, scientist,
inventor, philosopher, economist, printer, and musician
(he played the guitar, the harp, the viola da gamba,
and for good measure invented his own glass armonica),
had the talent to be creatively inventive when he was bored.
He must have been a clever doodler and droodler and riddler.
Once upon a time he constructed a rather beautiful $8\times8$
square, with lots of sums equal to 260. As he rather modestly
writes in his autobiography (published 1793),
``I was at length tired with sitting there to hear debates,
in which, as clerk, I could take no part, and which were
often so unentertaining that I was induc'd to amuse myself
with making magic squares or circles''. 

\begin{figure}[h]
\centering
\includegraphics[scale=0.55]{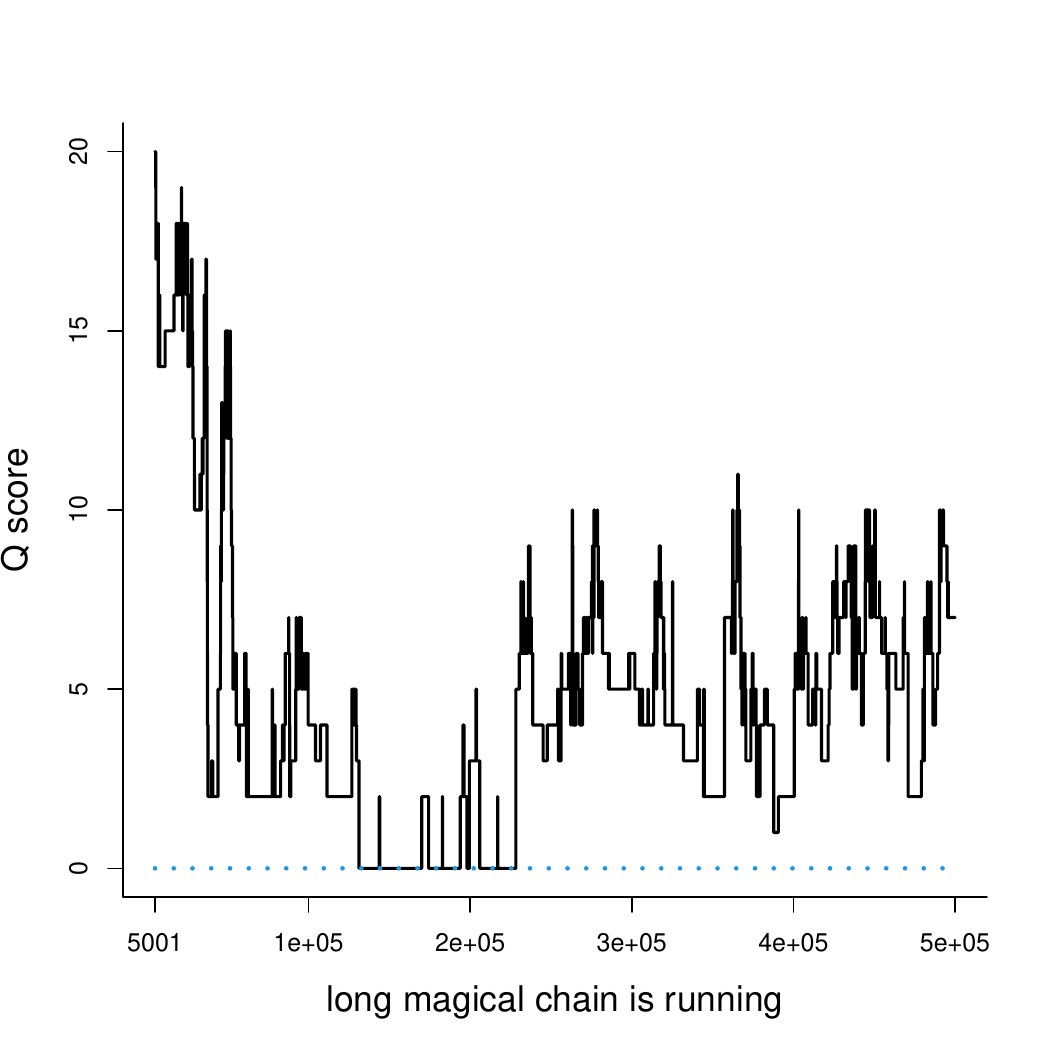}
\caption{Long magical square Markov chain to see when
  $Q(x)=R(x)+C(x)+D(x)+E(x)$ 
  hits zero, after which I can read off one of many
  solutions to the $8\times8$ magic square puzzle,
  with sums of rows, columns, diagonals all equal to 260,
  and the sum of the middle $4\times4$ block equal to 520. 
  In the figure I have discarded the first 5000 of
  the half a million random iterations.}
\label{figure:queen4}
\end{figure}

I'm not comparing myself to Franklin (I can't play
the viola da gamba, as yet, unfortunately), but the
approaches exhibited above, for sudoku solving, lend themselves
to clever searching for such $8\times8$ and similar
{\it magic squares}, with any set of desired properties.

Here's an example, where the tall order is to find a magic
$8\times8$ table, filled up with the numbers 1 to 64,
with all horizontal and vertical and diagonal sums equal
to 260, and at the same time demanding that also the
sum of the middle $4\times4$ block should be twice as big,
i.e.~520. I do this by setting up a probability model
$(1/k)\exp\{-\lambda Q(x)\}$, with
\beqn
Q(x)=R(x)+C(x)+D(x)+E(x), 
\eeqn 
with proper score functions for rows, columns, diagonals,
and the middle block, and then sampling away until
I spot a square with $Q(x)=0$. Specifically, with
\beqn
m=(1+\cdots+64)/8=\half\cdot64\cdot65/8=260
\eeqn
the required magic number for sums, and $B$ the mid-block
of size $4\times4$, 
\beqn
R(x)&=&\sum_{i=1}^8 \big|\sum_{j=1}^8 x[i,j]-m\big|, \\
C(x)&=&\sum_{j=1}^8 \big|\sum_{i=1}^8 x[i,j]-m\big|, \\
D(x)&=&\big|\sum_{i=1}^8 x[i,i]-m\big|+\big|\sum_{i=1}^8 x[i,9-i]-m\big|, \\
E(x)&=& \big|\sum_{(i,j)\in B} x[i,j]-2m\big|. 
\eeqn 
There are ridiculously many $64!=1.2688\cdot 10^{89}$ different
$8\times8$ squares $x$, using the numbers 1 to 64, and a ridiculously
tiny-tiny-tiny few of these are magic, with $R(x)=C(x)=D(x)+E(x)=0$. 
This is what I chanced to find, after half a million
or so of random iterations of my Markov chain,
see Figure \ref{figure:queen4}: 
\begin{large}
\begingroup
\setlength{\arraycolsep}{8pt}  
\beqn
\begin{bmatrix}
   39   &  38   &  13   &   8   &  27   &  62   &  47   &  26 \\
   10   &  54   &  51   &  53   &  22   &  19   &  37   &  14 \\
   36   &   2   &  12   &  48   &  61   &  35   &  20   &  46 \\
   55   &  30   &   1   &  18   &  31   &  57   &  52   &  16 \\
   56   &  43   &  29   &  64   &  15   &  44   &   5   &   4 \\
   49   &  40   &  33   &  34   &  21   &  17   &   7   &  59 \\
    6   &  25   &  63   &  11   &  42   &   3   &  60   &  50 \\
    9   &  28   &  58   &  24   &  41   &  23   &  32   &  45
\end{bmatrix} 
\eeqn
\endgroup 
\end{large} 

Also other types of magic squares can be found, via probability
models and long Markov chains. How long these Markov chains
need to be, before solutions are found, depend on both the
size and the wished-for constraints. Finding the $8\times8$
matrix above was after all relatively easy, in terms of R coding
and Markov iterations and laptop computing time (two minutes),
whereas the more demanding puzzle, to be described now, took
(i) just a bit more coding but (ii) ten times longer
chains and (iii) fifteen times more laptop time. The task
is as above, but also demanding that all five major
$4\times4$ blocks should have sums $2m=520$; the four
corner ones, and the one in the middle. I have no clear
idea how many solutions there might be, but my very long
Markov chain, using now the properly constructed 
\beqn
Q(x)=R(x)+C(x)+D(x)+E_1(x)+E_2(x)+E_3(x)+E_4(x)+E_5(x), 
\eeqn
found the following solution. A fair range of $\lambda$
values would work, and I did not attempt to optimise
that fine-tuning algorithmic parameter in any way,
but used $\lambda=0.444$.
\begin{large}
\beqn
\begingroup
\setlength{\arraycolsep}{6pt}  
\begin{bmatrix}
  27 & 50 & 6  & 48 & 43 & 39 & 46 &  1 \\
   3 & 56 & 60 &  8 & 14 & 49 & 61 &  9 \\
  11 & 28 & 45 & 53 & 24 & 40 & 25 & 34 \\
  35 & 42 &  7 & 41 & 16 & 36 & 19 & 64 \\
  22 & 17 & 44 & 26 & 54 & 13 & 21 & 63 \\
  59 &  5 & 33 & 57 & 29 &  2 & 37 & 38 \\
  52 & 32 & 55 & 12 & 62 & 23 &  4 & 20 \\
  51 & 30 & 10 & 15 & 18 & 58 & 47 & 31 
\end{bmatrix}
\endgroup 
\eeqn
\end{large} 

Since I could fiddle with my R code and make it work,
and since I was sufficiently curious, I also attacked
a grander problem, and trust Benjamin Franklin
would have been impressed. The taller task is to find
a $10\times10$ matrix, with sums of rows, of columns,
of the two diagonals, and also of all ten canonical 
sub-blocks of size $2\times5$, all equal to the magical
number $m=505$. After a few million steps and perhaps
twenty minutes of laptop computing time my chain
of random $10\times10$ matrices hit this solution: 
\begin{large}
\beqn
\begingroup
\setlength{\arraycolsep}{6pt}  
\begin{bmatrix}
36 & 39 & 48 & 91 & 64 &  9 & 49 & 34 & 52 & 83 \\
92 & 17 &  4 & 42 & 72 & 94 & 57 & 28 & 86 & 13 \\
88 & 21 & 85 & 38 & 51 & 54 & 61 & 65 & 32 & 10 \\
79 & 73 & 37 & 18 & 15 & 90 & 23 & 87 & 80 &  3 \\
19 &  8 & 63 & 71 & 99 &  5 & 40 & 69 & 31 & 100 \\
68 & 35 & 96 & 45 &  1 & 84 & 11 & 25 & 74 & 66 \\
33 & 82 & 44 & 95 & 67 & 22 & 53 & 27 & 12 & 70 \\
 2 & 55 & 62 & 59 &  6 & 47 & 58 & 43 & 98 & 75 \\
81 & 78 & 46 & 30 & 41 & 76 & 60 & 50 & 14 & 29 \\
 7 & 97 & 20 & 16 & 89 & 24 & 93 & 77 & 26 & 56
\end{bmatrix}
\endgroup 
\eeqn
\end{large} 
You may indeed check that
$R(x)=0$ (ten rows ok),
$C(x)=0$ (ten columns ok),
$D(x)=0$ (both diagonals ok),
$B(x)=0$ (the ten $2\times5$ blocks ok). 

The book Dickter (2014) gives a fascinating account
of the rich cultural-historical landscape of magic squares.
It would have surprised the constructors of such puzzles,
over many centuries, 
to learn that these can be solved via probability theory 
and Markov chains. 

\def\blocktolist{{\tt blocktolist}}
\def\listtoblock{{\tt listtoblock}} 

\section{A few remarks} 

\noindent 
{\bf A.}
A.A. Markov (1856--1922) invented Markov chains in about 1910,
and the world's first serious data analysis using Markov chains
was carried out by Markov himself, in 1913. Amazingly,
he went through the first 20,000 letters of Pushkin's
epic poem {{\onegin}},  
keeping track of how vowels
and consonants followed consonants and vowels, the point
being to clearly demonstrate that the vowel-consonant gender
of the letters of words are not statistically independent.
See more discussion on this in N.L. Hjort and C. Varin's
paper {\it ML, PL, QL in Markov chain models}
(Scandinavian Journal of Statistics, 2008),
where we also study longer-memory processes. So Markov's
sample space, for that application, had two elements.
I imagine that he would have been delighted to see how
I can use his {\markovchains}, 
in enormously big sample spaces, to read off solutions
to highly complicated puzzles, like sudoku and magic squares.

\hopp
{\bf B.}
There would be several ways to improve my R code, both algorithmically
and mathematically. To do the book-keeping I needed to invent
$\blocktolist(x)$ and $\listtoblock(x)$ 
functions, and I do the $Q(x_\prop)-Q(x_\prev)$
computations by straightforward brute force. It is clear
that if a puzzler sees some clear numbers to fill
in for the blanks, those can be plugged in, simplifying
not the algorithm per se but the speed with which it lands
at zero.

Yet other number theoretic based tricks could
be built into the algorithm and my R code,
since some of the choices made in the $x_\prop$ step 
are sillier than others, but the relative statistical beauty
of the Nils Sudoku Solver$^{\rm TM}$ is its simplicity
and generality. In particular, the code doesn't know,
and doesn't care, whether the sudoku puzzle is seen as `hard'
or `relatively easy'.

\hopp
{\bf C.} The role of the $\lambda$ parameter is partly
to dictate the sharpness of the probability distribution around
its peak, the $x^*$ solution.
It sounds good to allow a somewhat large $\lambda$
since that means that once you're near the peak, you'll
never slide down again. At the same time one needs $\lambda$ 
to be small enough to allow the chain to jump around
a bit for a while. There's a possibility of getting trapped
close to a fine-looking mountain, which indeed is tall
(perhaps $Q(x)$ is as low as 8 or 6 or 4 or 2),
but not the same as the Very Tallest One, the sudoko solution $x^*$. 
Perhaps the sudoku chain is labouring away, for a while,
climbing K2 or Kangchenjunga or Lhotse, but eventually
it should discover that there's a Mount Everest elsewhere,
and give itself the chance to backtrack and climb upwards again.
For these reasons $\lambda$ should be somewhat small.

\medskip 
\begin{figure}[h]
\centering
\includegraphics[scale=1.11]{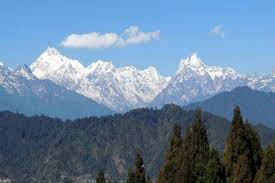}
\caption{The sudoku Markov chain will eventually
  find the tallest peak. My Himalayan metaphor is misleading
  regarding the size of the problem; the sample space in which
  the chain selflessly struggles, with a hundred thousand
  attempts per minute, is astronomical rather than merely planetary big.}
\label{figure:queen5}
\end{figure}


There are mathematical theorems securing that the random chain will
succeed in finding the Tallest Peak in the World,
with high enough probability, by having $\lambda$ 
start small and then very gradually increase, but it is
difficult to translate such theorems into very clear guidelines.
Based on having solved just a few sudoko puzzles,
\fermat{Four Color \#318, 1951}I think
the Tryle N.~Error attitude is fine, perhaps with $\lambda=1$ 
as an ok start value (with the $Q(x)$ function used above).

\hopp
{\bf D.}
The $f(x)=(1/k)\exp\{-\lambda Q(x)\}$ model has been constructed
to favour good-looking sudoku tables, not far off from the
final solution. I'm free to also consider negative $\lambda$ 
values, however, which seriously turns the tables, suddenly
favouring high $Q(x)$ values rather than low ones.
The Markov chain apparatus still obeys, and can churn out
clearly horrible-looking sudoku tables, as far away from
the genuine solution as possible, in its still enormous
space of possibilities. Here's one such, generated by
letting the chain roll on for a while, with a negative model
tuning parameter. It's still a sudoku table, with 1-to-9
in each $3\times3$ block, but the rows and columns of the $9\times9$
table are nine far cries away from being what the puzzler wants:

\begin{large}
\beqn
\begingroup
\setlength{\arraycolsep}{8pt}  
\begin{matrix}
\begin{bmatrix}
  5 & 3 & 1 \\
  6 & 4 & 2 \\
  7 & 9 & 8
\end{bmatrix} \quad 
\begin{bmatrix}
  2 & 7 & 6 \\
  1 & 9 & 5 \\
  4 & 3 & 8
\end{bmatrix} \quad
\begin{bmatrix}
  1 & 5 & 2 \\
  3 & 9 & 4 \\
  8 & 6 & 7
\end{bmatrix} \\ \\[1pt] 
\begin{bmatrix}
  8 & 9 & 3 \\
  4 & 6 & 1 \\
  7 & 5 & 2
\end{bmatrix} \quad
\begin{bmatrix}
  7 & 6 & 9 \\
  8 & 1 & 3 \\
  4 & 2 & 5
\end{bmatrix} \quad
\begin{bmatrix}
  7 & 9 & 3 \\
  2 & 8 & 1 \\
  4 & 5 & 6
\end{bmatrix} \\ \\[1pt]
\begin{bmatrix}
  8 & 6 & 1 \\
  5 & 3 & 2 \\
  7 & 9 & 4
\end{bmatrix} \quad
\begin{bmatrix}
  3 & 2 & 6 \\
  4 & 1 & 9 \\
  5 & 8 & 7
\end{bmatrix} \quad
\begin{bmatrix}
  2 & 8 & 3 \\
  1 & 6 & 5 \\
  4 & 7 & 9
\end{bmatrix}
\end{matrix}
\endgroup 
\eeqn 
\end{large}

\hopp
{\bf E.}
Not surprisingly, there are various mathematically based non-random
algorithms, which work for classes of sudoku puzzles (try the net).
I have not attempted to look into these, and neither have
I tried to compare my Markov chains with any of those.
I expect that when a non-random algorithm works, then it will
work more efficiently than my after all simple Markov chains.
But my construction is pleasantly clear (well, in principle,
give and take some plundering work to get the R code in shape),
is fun to watch in action, and working through a few million
steps is easier than attempting to search more brutally through
bigger parts of the $10^{23}$ possible cases.

\hopp
{\bf F.}
This article is more probabilistical than
statistical in nature. I have not constructed or fitted models
to data, but invented probability models to help me search for
the mini-needle in the tera-haystack. Yet there are important
application areas, with complex and massive data,
like images and texts, where models of the type
\beqn
f(x)=(1/k)\exp\{-\lambda^\tr Q(x)\}
    =(1/k)\exp\Bigl\{-\sum_{j=1}^p \lambda_j Q_j(x)\Bigr\}
\eeqn 
are useful, typically with a multidimensional $Q(x)=(Q_1(x),\ldots,Q_p(x))$
and hence statistical model parameters $\lambda=(\lambda_1,\ldots,\lambda_p)$,
say. If you see photos of seven thousand men, or dogs, models could
be built of this type, with specially designed $Q_j(x)$
functions, in order to look for differences, or signs of homogeneity,
or to analyse the levels of variation. Such applications would
then involve estimating and assessing the $\lambda$ 
parameters, via frequentist or Bayesian setups.
Tackling the issues would again involve running Markov chains.
Situations and models of this type may be found in
Claeskens and Hjort (2008), Schweder and Hjort (2016),
Hjort and Stoltenberg (2026). 

\hoppp
{\bf Thanks.} 
I am often enough grateful to C\'eline Cunen for valuable
discussions, also here, on aspects taken up in this article. 

\section*{References}

\begin{small}
\parindent0pt
\parskip4pt

Claeskens, G. and Hjort, N.L. (2008).
{\it Model Selection and Model Averaging.}
Cambridge University Press.

Dickter, R. (2014).
{\sl Number Time Archetype: The Significance of Magic Squares
in China and the West.} 
Grass Valley, Grass Valley.

Edwards, A.W.F. (2025).
{\it The Latin Square: Essays in Defence of R.A.~Fisher.}
Cam Rivers Publishing, Cambridge, UK. 

Franklin, B. (1791). 
{\it The Autobiography of Benjamin Franklin.}
M\'emoires de la vie priv\'ee de Benjamin Franklin,
Buisson, Paris; then published in English, by J.~Parson's, London, 1793. 

Gei\ss ler, A. (1878).
Über die Phantasmen während des Einschlafens.
{\it Philosophische Studien}, vol.~1, 83–93.

Hjort, N.L. (2019a).
Your Mother is Alive with Probability One Half.
{\it FocuStat Blog Post.}

Hjort, N.L. (2019b).
The Magic Square of 33. 
{\it FocuStat Blog Post.}

Hjort, N.L. and Varin, C. (2008).
ML, PL, QL in Markov chain models. 
{\it Scandinavian Journal of Statistics}, {\bf 35}, 64--82. 

Hjort, N.L. and Stoltenberg, E.Aa. (2026).
{\it Statistical Inference: 600 Exercises and 100 Stories.}
Cambridge University Press. 

Schweder, T. and Hjort, N.L. (2016).
 {\it Confidence, Likelihood, Probability:
   Statistical Inference With Confidence Distributions.}
 Cambridge University Press, Cambridge.

\end{small}

\end{document}